\documentclass[american,english,aps,prb,preprint,showpacs]{revtex4}
\usepackage[T1]{fontenc}
\usepackage[latin9]{inputenc}
\usepackage{amsmath}
\usepackage{graphicx}
\usepackage{esint}

\makeatletter
\@ifundefined{textcolor}{}
{%
 \definecolor{BLACK}{gray}{0}
 \definecolor{WHITE}{gray}{1}
 \definecolor{RED}{rgb}{1,0,0}
 \definecolor{GREEN}{rgb}{0,1,0}
 \definecolor{BLUE}{rgb}{0,0,1}
 \definecolor{CYAN}{cmyk}{1,0,0,0}
 \definecolor{MAGENTA}{cmyk}{0,1,0,0}
 \definecolor{YELLOW}{cmyk}{0,0,1,0}
 }

\makeatother

\usepackage{babel}

\begin{document}

\title{Doubly Quantized Vortices in Bulk Ginzburg-Landau Superconductors}

\author{Mark C. Sweeney}

\author{Martin P. Gelfand}

\email{martin.gelfand@colostate.edu}

\address{Department of Physics, Colorado State University, Fort Collins, Colorado
80523-1875 USA}
\begin{abstract}
We have extended Brandt's method for accurate, efficient calculations
within Ginzburg-Landau theory for periodic vortex lattices at arbitrary
mean induction to lattices of {}``doubly quantized'' vortices. 
\end{abstract}
\pacs{74.25.Uv, 74.20.De}
\maketitle

\section{\label{sec:Introduction}Introduction}

In bulk type-II superconductors, mixed states in which the vortices
carry more than one flux quantum are unstable with respect to the
usual vortices that carry a single flux quantum. Near the upper critical
field this result was established by Abrikosov,\cite{abrikosov_magnetic_1957}
while near the lower critical field it is implied by Matricon's\cite{matricon_thesis}
calculations for isolated vortices. As far as we are aware there have
been no explicit calculations within Ginzburg-Landau (GL) theory for
lattices of multiply quantized vortices at magnetic field values between
these two limits; but since there is no reason to believe that such
lattices should be stable at intermediate fields there has been no
reason to carry out the calculations.

Our motivation for doing so comes from considering thin films of type-I
superconductors. In a sufficiently weak perpendicular field the magnetic
flux penetrates the sample in vortices rather than intermediate state
structures, and in some circumstances multiply quantized vortices
are present in Ginzburg-Landau theory calculations\cite{lasher_mixed_1967,callaway_magnetic_1992}
and experiments.\cite{hasegawa_magnetic-flux_1991} We have been interested\cite{sweeney_simple_2010}
in working out the phase diagram for thin films of type-I GL superconductors
in arbitrary magnetic field, following Brandt's 
approach\cite{brandt_ginsburg-landau_1972,brandt_precision_1997}
to accurate and efficient solutions of the GL equations. This approach
involves iterative solutions of equations, and good initial values
are essential because convergence is not guaranteed. It turns out
that a solution of the GL equations in bulk can be used as starting
point for the film problem---this is the case even for type-I superconductors,
where the bulk vortex lattice solutions are unstable with respect
to the normal state. Hence, calculations for lattices of multiply
quantized vortices in bulk superconductors are almost a prerequisite
for the more physically motivated calculations in films.

For brevity, in this paper {}``singles'' and {}``doubles'' are
synonymous with lattices of singly and doubly quantized vortices (carrying
one and two flux quanta, respectively).

The formal developments we present are an extension of Brandt's work
in Ref.~\onlinecite{brandt_precision_1997}. In Sec.~\ref{sec:Form-of-the-GLSolutions}
we describe how Brandt's Ansatz and iteration scheme for singles
are modified for doubles. In Sec.~\ref{sec:Results} we offer illustrative results
for order parameter and magnetic induction profiles, as well as the Gibbs free energy,
for doubles and singles in a type-II superconductor.
In the Appendix
we discuss the solutions of the linearized GL equations, which serve
as initial values for the iterative calculations.

\section{\label{sec:Form-of-the-GLSolutions}Formalism and Implementation}

\subsection{Form of solutions for doubles}

Consider a vortex in which the phase of the order parameter changes
by $2\pi p$ on circling the vortex core. If that core is at the origin,
then the order parameter behaves as $\psi\sim r^{p}e^{ip\theta}$
as $r\to0$ (see for example Tinkham\cite{tinkham_introduction_1996}
Sec.~5.1 ) and the modulus squared order parameter as $\omega\equiv\left|\psi\right|^{2}\sim r^{2p}$.
In this paper we focus on the $p=2$ case. Brandt suggests\cite{brandt_precision_1997}
that for a lattice with one vortex per primitive cell we adopt the
Ansatz

\begin{equation}
\omega\left(\mathbf{r}\right)=\sum_{\mathbf{K}}a_{\mathbf{K}}\left[1-\cos\left(\mathbf{K}\cdot\mathbf{r}\right)\right]^{2}\label{eq:omegaansatz1}\end{equation}
in which $\mathbf{r}$ is a two-dimensional vector and $\mathbf{K}$
runs over reciprocal lattice vectors excluding the origin (as will
be the case for all sums over $\mathbf{K}$ henceforth). This form
satisfies the requirements of periodicity and fourth-power behavior
near vortex cores. It turns out to be useful to express this with
only first powers of cosines, 

\begin{equation}
\omega\left(\mathbf{r}\right)=\sum_{\mathbf{K}}a_{\mathbf{K}}\left[\frac{3}{2}-2\cos\left(\mathbf{K}\cdot\mathbf{r}\right)+\frac{1}{2}\cos\left(2\mathbf{K}\cdot\mathbf{r}\right)\right]\label{eq:omegaansatz2}\end{equation}

In a bulk superconductor $\mathbf{B}\left(\mathbf{r}\right)=B\left(\mathbf{r}\right)\mathbf{\hat{z}}$.
For small $r$ the induction satisfies $B\left(\mathbf{r}\right)\approx B\left(0\right)-\frac{1}{2\kappa}\omega\left(\mathbf{r}\right)$,
so $B\left(0\right)-B\left(r\right)\sim r^{4}$. The small-$r$ behavior
suggests the following form for the deviation from mean induction,
$b\left(\mathbf{r}\right)=B\left(\mathbf{r}\right)-\bar{B}$, \begin{equation}
b\left(\mathbf{r}\right)=\sum_{\mathbf{K}}b_{\mathbf{K}}\left[2\cos\left(\mathbf{K}\cdot\mathbf{r}\right)-\frac{1}{2}\cos\left(2\mathbf{K}\cdot\mathbf{r}\right)\right]\label{eq:bansatz}\end{equation}

The supervelocity can be decomposed as $\mathbf{Q}\left(\mathbf{r}\right)=\mathbf{Q}_{A}\left(\mathbf{r}\right)+\mathbf{q}\left(\mathbf{r}\right)$,
where $\mathbf{Q}_{A}$ is the supervelocity in the Abrikosov limit,
satisfying $\nabla\times\mathbf{Q}_{A}\left(\mathbf{r}\right)=\left[\bar{B}-2\Phi_{0}\delta\left(\mathbf{r}\right)\right]\mathbf{\hat{z}}$,
(with $\Phi_{0}$ the flux quantum) and the deviation from the Abrikosov
form has the property that $\hat{\mathbf{z}}\cdot(\nabla\times\mathbf{q}\left(\mathbf{r}\right))=b\left(\mathbf{r}\right)$.
This last relation implies

\begin{equation}
\mathbf{q}\left(\mathbf{r}\right)=\sum_{\mathbf{K}}b_{\mathbf{K}}\frac{\mathbf{\hat{z}}\mathbf{\times K}}{\left|\mathbf{K}^{2}\right|}\left[2\sin\left(\mathbf{K}\cdot\mathbf{r}\right)-\frac{1}{4}\sin\left(2\mathbf{K}\cdot\mathbf{r}\right)\right]\label{eq:qansatz}\end{equation}

The mean induction $\bar{B}$ fixes the area $S$ of the lattice unit
cell, through $S=2\Phi_{0}/\bar{B}$. The unit cell has primitive
lattice vectors $\mathbf{R}_{10}=x_{1}\hat{\mathbf{x}}$, $\mathbf{R}_{01}=x_{2}\hat{\mathbf{x}}+y_{2}\hat{\mathbf{y}}$
and in those terms $S=(\mathbf{R}_{10}\times\mathbf{R}_{01})\cdot\hat{\mathbf{z}}=x_{1}y_{2}$.
The general reciprocal lattice vector is $\mathbf{K}_{mn}=2\pi\left[my_{2}\hat{\mathbf{x}}+\left(mx_{2}+nx_{1}\right)\hat{\mathbf{y}}\right]/S$.

\subsection{\label{sec:GL-Equations}Iterative expressions}

The GL free energy per unit volume $F$, referenced from the Meissner
state, may be expressed in gauge invariant form and in the standard
reduced units as\cite{brandt_precision_1997}

\begin{equation}
F=\left\langle \frac{1}{2}-\omega+\frac{1}{2}\omega^{2}+\frac{\left|\nabla\omega\right|^{2}}{4\kappa^{2}\omega}+\omega\left|\mathbf{Q}\right|^{2}+\left|\mathbf{B}\right|^{2}\right\rangle \label{eq:FreeEn2}\end{equation}
where angle brackets denote integration over a two-dimensional unit
cell, $\left\langle \cdots\right\rangle =\frac{1}{S}\int_{S}\cdots dx\, dy$.
Extremalization of $F$ yields the first GL equation \begin{equation}
-\nabla^{2}\omega=2\kappa^{2}\left[\omega-\omega^{2}-\omega\left|\mathbf{Q}\right|^{2}-g\right]\label{eq:GLEQ1}\end{equation}
where\foreignlanguage{american}{\begin{equation}
g\equiv|\nabla\omega|^{2}/4\kappa^{2}\omega\label{eq:gdef}\end{equation}
}and the second GL equation 

\begin{equation}
\nabla\times B\hat{\mathbf{z}}=-\omega\mathbf{Q}\label{eq:GLEQ2}\end{equation}

The first GL equation leads to an iterative equation for $a_{\mathbf{K}}$.
Following Brandt, a stabilizing term $2\kappa^{2}\omega$ is added
to both sides of Eq.~\eqref{eq:GLEQ1} and then the equation is multiplied
by $\cos\mathbf{K}\cdot\mathbf{r}$ and integrated over the unit cell.
The expansion \eqref{eq:omegaansatz2} is inserted for $\omega$ on
the left side of the equation, and the orthogonality relation $\left\langle \cos\left(\mathbf{K}\cdot\mathbf{r}\right)\cos\left(\mathbf{K}'\cdot\mathbf{r}\right)\right\rangle =\frac{1}{2}\delta_{\mathbf{K},\mathbf{K}'}$
enables the integral on the left side to be done analytically. Rearranging
leads to an identity which we treat as a step in an iterative solution
for $a_{\mathbf{K}}$,

\begin{equation}
\begin{aligned}a_{\mathbf{K}}:=\frac{2\kappa^{2}}{\left|\mathbf{K}\right|^{2}+2\kappa^{2}}\langle(-2\omega+\omega^{2}+\omega\left|\mathbf{Q}\right|^{2}+g)\cos\mathbf{K}\cdot\mathbf{r}\rangle\\
+\frac{1}{4}a_{\mathbf{K}/2}\end{aligned}
\label{eq:aKupdate}\end{equation}
If $\mathbf{K}/2$ is not a reciprocal lattice vector then $a_{\mathbf{K}/2}\equiv0$;
we will refer to such reciprocal lattice vectors as {}``fundamentals.''
Eq.~\eqref{eq:aKupdate} should be compared with the corresponding
relation for singles, Eq.~(11) in Ref.~\onlinecite{brandt_precision_1997}:
the only differences are the existence of the second term and a factor
of two in the first term.

The next step in the iterative scheme is to rescale all of the $a_{\mathbf{K}}$
so as to minimize $F$. This goes through without modification \begin{equation}
a_{\mathbf{K}}:=a_{\mathbf{K}}\langle\omega-\omega\left|\mathbf{Q}\right|^{2}-g\rangle/\langle\omega^{2}\rangle\label{eq:aKupdate2}\end{equation}

The iterative equation for $b_{\mathbf{K}}$ is derived in the same
manner as Eq.~\eqref{eq:aKupdate}. Taking the curl of Eq.~\eqref{eq:GLEQ2},
adding a stabilizing term $-\langle\omega\rangle B\left(\mathbf{r}\right)$
to both sides, multiplying by $\cos\mathbf{K}\cdot\mathbf{r}$, integrating
over the unit cell, inserting the expansion \eqref{eq:bansatz} on
the left hand side, applying orthogonality of cosines, and rearranging
leads to \begin{equation}
\begin{aligned}b_{\mathbf{K}}:=-\frac{\langle[(\omega-\langle\omega\rangle)B\left(\mathbf{r}\right)+(\nabla\omega\times\mathbf{Q})\cdot\hat{z}]\cos\mathbf{K}\cdot\mathbf{r}\rangle}{\left|\mathbf{K}\right|^{2}+\langle\omega\rangle}\\
+\frac{1}{4}b_{\mathbf{K}/2}\end{aligned}
\label{eq:bKupdate}\end{equation}
Again, if $\mathbf{K}$ is a fundamental then the coefficient $b_{\mathbf{K}/2}\equiv0$.

The GL equations are solved, in principle, by cycling through Eqs.~\eqref{eq:aKupdate},
\eqref{eq:aKupdate2}, and \eqref{eq:bKupdate} until the coefficients
converge to the desired level of precision. In practice we find that
the equations as written do not usually converge to a physical solution;
however, by {}``mixing'' the $a_{\mathbf{K}}$ that comes out of
\eqref{eq:aKupdate} with the value from the prior iteration (and
likewise for the $b_{\mathbf{K}}$ produced by \eqref{eq:bKupdate})
the convergence of the algorithm is much improved, though at the cost
of more iterations. We have not attempted to determine optimal mixing
parameters. Taking 90\% of the prior iteration plus 10\% of the current
iteration is sufficient for every calculation we have carried out
so far.

Even with mixing, it is crucial to have a good initial guess for the
$a_{\mathbf{K}}$ and $b_{\mathbf{K}}$. Brandt has demonstrated\cite{brandt_precision_1997}
that the solution of the linearized GL equations for the $a_{\mathbf{K}}$,
together with $b_{\mathbf{K}}=0$ for all $\mathbf{K}$, serves well
for the initial values for singles, and we find the same to be true
for doubles. Constructing solutions to the linearized GL equations
for doubles in terms of the $a_{\mathbf{K}}$ is not trivial, and
we detail our method in Appendix~\ref{sec:AppendixA}. The solution
to the linearized GL equations is used to begin the iteration cycle
in Eq.~\eqref{eq:aKupdate2}.

\subsection{Implementation issues\label{sub:Implementation-issues}}

In order to have a finite computational problem the expansions for
$\omega$, $b$, and $\mathbf{q}$ must be truncated; and the iterations
involve integrals over the unit cell which must be numerically evaluated.
These two issues are related. For singles, it is sufficient to carry
out the quadrature by summation of values on a grid aligned with the
primitive lattice vectors, and to include in the expansions only $|\mathbf{K}|\leq K_{\mathrm{max}}$
with $K_{\mathrm{max}}$ chosen so that the number of reciprocal lattice
vectors is the same as the number of points in the integration grid.
For doubles the situation is more complicated.

Observe that the expansion \eqref{eq:omegaansatz2} for $\omega$
can be rearranged so that it has nearly the same form as for singles,
\begin{equation}
\omega\left(\mathbf{r}\right)=\sum_{\mathbf{K}}\left[2a_{\mathbf{K}}-\frac{1}{2}a_{\mathbf{K}/2}\right]\left[1-\cos\left(\mathbf{K}\cdot\mathbf{r}\right)\right]\label{eq:OmegaAnsatz3}\end{equation}
When $\mathbf{K}$ is a fundamental, as defined following Eq.~\eqref{eq:aKupdate},
$a_{\mathbf{K}/2}\equiv0$. Eqs.~\eqref{eq:omegaansatz1}, \eqref{eq:omegaansatz2}
and \eqref{eq:OmegaAnsatz3} are identical for infinite sums, but
they are different when truncated. As discussed in Appendix~\ref{sec:AppendixA},
the $a_{\mathbf{K}}$ that solve the linearized GL equations for doubles
do not fall off in a Gaussian manner like they do for singles; however,
the $2a_{\mathbf{K}}-\frac{1}{2}a_{\mathbf{K}/2}$ are approximately
Gaussian. This motivates the following truncation scheme for constructing
$\omega$ when evaluating the integrals over the unit cell: use Eq.~\ref{eq:OmegaAnsatz3},
including in the sum reciprocal lattice vectors with $|\mathbf{K}|\leq K_{\mathrm{max}}$
except for fundamentals with $K_{\mathrm{max}}/2<|\mathbf{K}|\leq K_{\mathrm{max}}$.
Expressions analogous to Eq.~\ref{eq:OmegaAnsatz3} exist for $b$
and $\mathbf{q}$, namely\begin{equation}
b\left(\mathbf{r}\right)=\sum_{\mathbf{K}}\left[2b_{\mathbf{K}}-\frac{1}{2}b_{\mathbf{K}/2}\right]\cos\left(\mathbf{K}\cdot\mathbf{r}\right)\label{eq:bansatz2}\end{equation}
\begin{equation}
\mathbf{q}\left(\mathbf{r}\right)=\sum_{\mathbf{K}}\left[2b_{\mathbf{K}}-\frac{1}{2}b_{\mathbf{K}/2}\right]\sin\left(\mathbf{K}\cdot\mathbf{r}\right)\frac{\mathbf{\hat{z}}\mathbf{\times K}}{\left|\mathbf{K}^{2}\right|}\end{equation}
and we apply the same truncation scheme. 

We conclude this section with a warning. It is tempting to define
$c_{\mathbf{K}}\equiv2a_{\mathbf{K}}-\frac{1}{2}a_{\mathbf{K}/2}$,
$d_{\mathbf{K}}\equiv2b_{\mathbf{K}}-\frac{1}{2}b_{\mathbf{K}/2}$,
and carry out iterative calculations for those quantities, by moving
the $\frac{1}{4}a_{\mathbf{K}/2}$ from the right side of Eq.~\eqref{eq:aKupdate}
to the left (and likewise for Eq.~\eqref{eq:bKupdate}): the resulting
equations have exactly the form of Eqs.~11 and 13 from Ref.~\onlinecite{brandt_precision_1997}
for the iterations of the coefficients for singles. Doing this invariably
leads to a {}``singles-like'' solution ($\omega\sim r^{2}$ and
$B\left(0\right)-B\left(r\right)\sim r^{2}$ for small $r$) which
is inconsistent with the assumed forms of $\bar{B}$, $\mathbf{Q}_{A}$
and $S$; in addition this unphysical solution is a free energy saddle
point rather than a minimum.

\section{\label{sec:Results}Results and Conclusions}

To illustrate the effectiveness of the calculational scheme described
in the previous section, we will present results for $\kappa=1$ for
triangular arrays of both doubly and singly quantized vortices. In
comparing doubles with singles it is important to do the calculations
on an equal footing; with this in mind, in order to have the same
spacing between real-space grid points in both calculations we use
about twice as many grid points for doubles than for singles. For
$\bar{B}$ greater than $\mu_{0}H_{c2}/5$ we typically use 32 points
along each primitive lattice vector for singles and 46 for doubles.
Singles and doubles then have the same $K_{\mathrm{max}}$ but the
doubles calculation includes twice as many reciprocal lattice vectors
as the singles calculation. A typical singles calculation converges
in approximately 50 iterations while the doubles require about four
times as many. At lower inductions the area of the vortex core becomes
considerably less than the area of the unit cell, and in order to
represent the solutions well the real-space sampling needs to be refined
by increasing the number of grid points per unit cell edge to 64 or
96 for singles and with proportionally increased numbers for doubles.
The growth in the number of grid points and reciprocal lattice vectors
puts a practical lower limit on the mean induction of $\mu_{0}H_{c2}/10$.

\begin{figure}
\includegraphics[width=1\columnwidth]{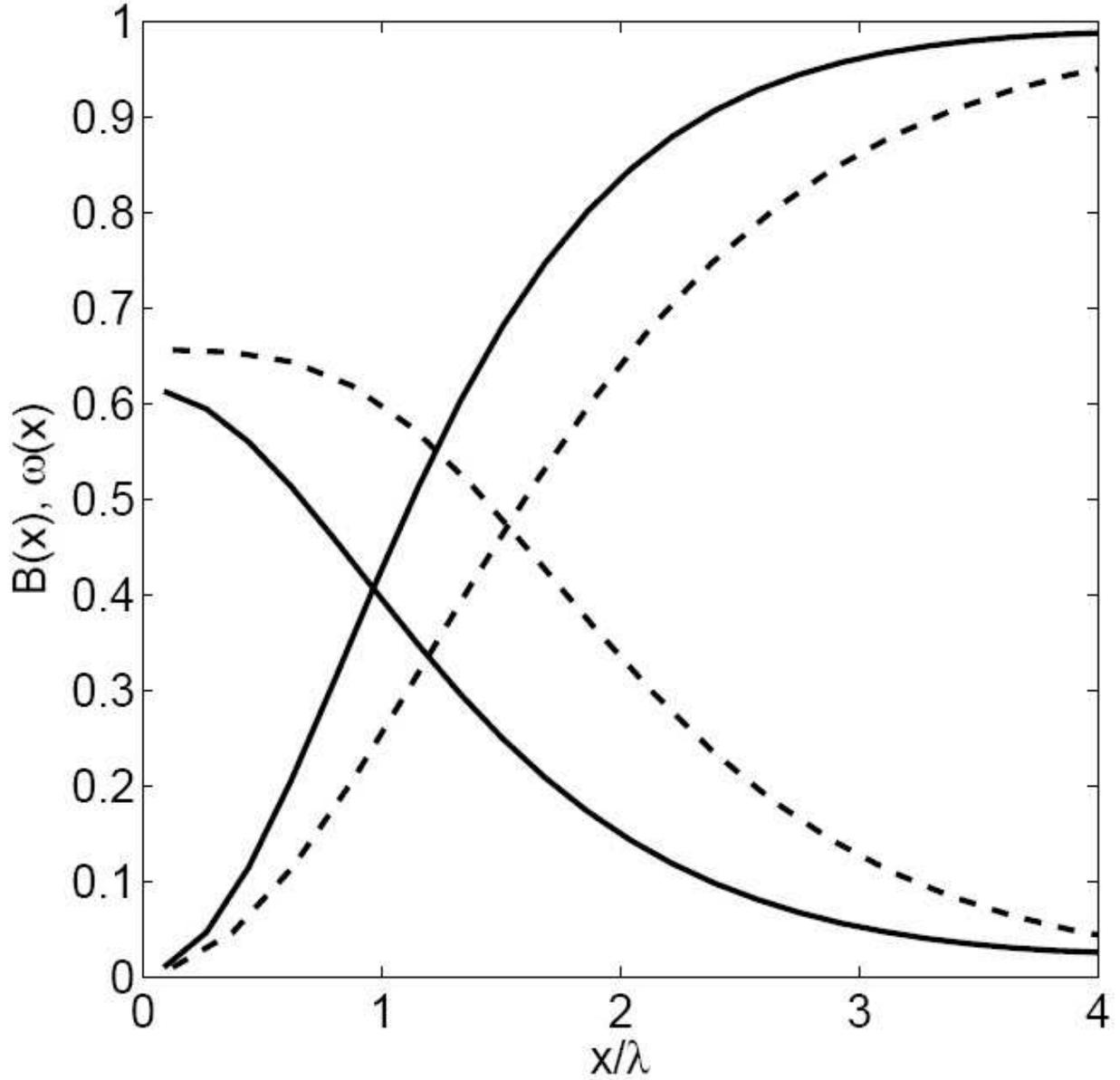}\caption{\label{fig:CrossSections}Cross sections along $y=0$ of induction
(curves decreasing from $x=0$) and squared order parameter (curves
increasing from $x=0$) for doubly quantized (dashed) and singly quantized
(solid) triangular vortex lattices at the same parameters $\kappa=1$
and $\bar{B}=\mu_{0}H_{c2}/10$. The inter-vortex spacing for singles
is $8.5\lambda$ and is greater by a factor of $\sqrt{2}$ for doubles.}

\end{figure}

\begin{figure}
\includegraphics[width=1\columnwidth]{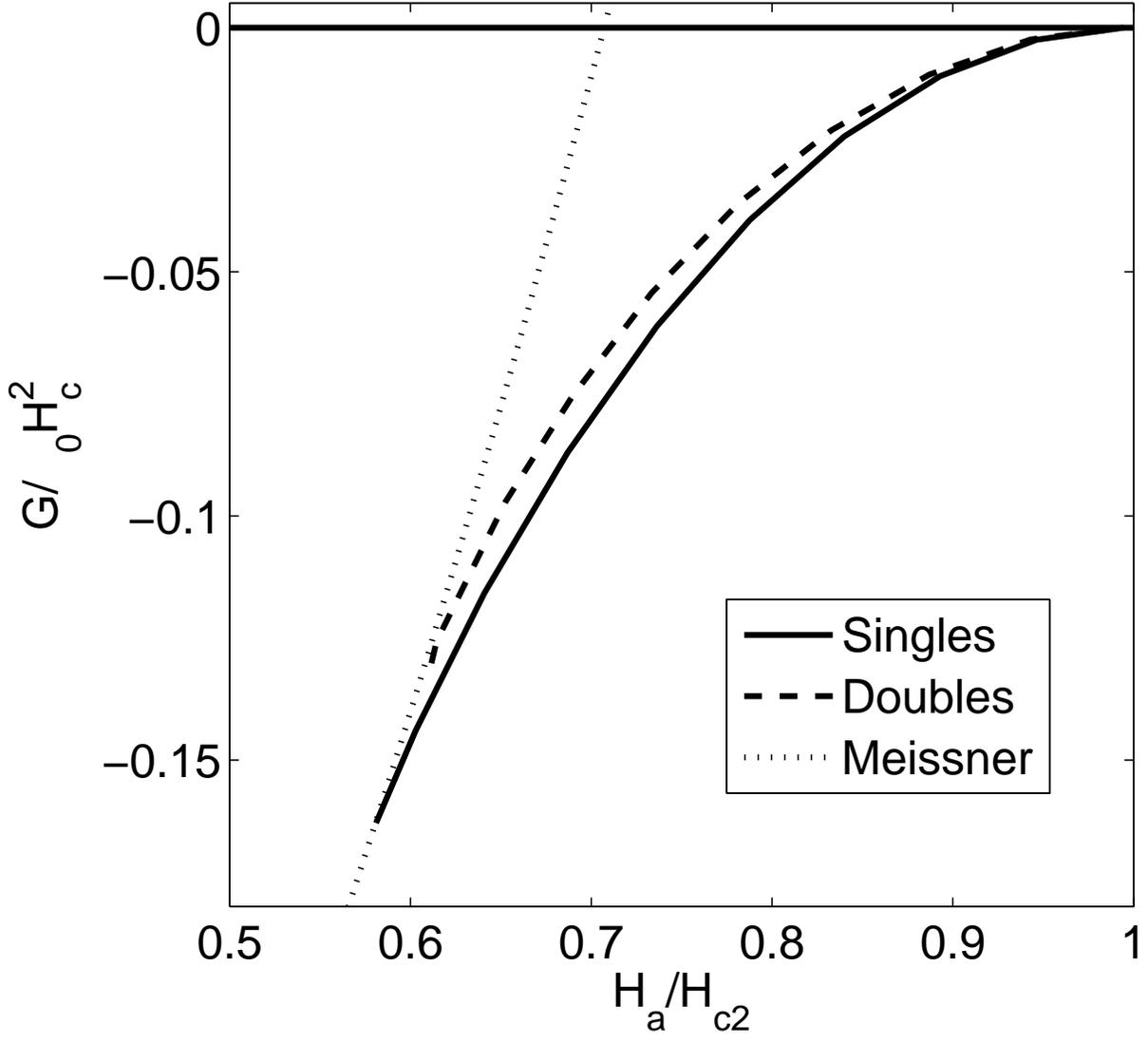}\caption{\label{fig:Gibbs}Gibbs free energy density, referenced from the normal
state, as a function of applied field at GL parameter $\kappa=1$. }

\end{figure}

In Figure~\ref{fig:CrossSections} we show the order parameter and
induction along a line connecting two adjacent vortices at $\bar{B}=\mu_{0}H_{c2}/10$
for both doubles and singles. As one would expect, the vortex cores
for doubles are wider than for singles. In Fig.~\ref{fig:Gibbs}
we present the Gibbs free energy density $G=F-\mathbf{H}_{a}\cdot\mathbf{B}$
in the full range of applied fields $H_{c1}\leq H_{a}\leq H_{c2}$
for triangular singles and doubles. The applied field $H_{a}$ is
calculated in the same manner as in Refs.~\onlinecite{klein_virial_1991,brandt_precision_1997},
based on the virial theorem of Doria, Gubernatis, and Rainer.\cite{doria_virial_1989}
These results confirm that doubles are thermodynamically unstable
in bulk type-II superconductors. Similar results are obtained for
other values of $\kappa>1/\sqrt{2}$. The calculations also yield
convergent results for type-I superconductors but in such cases all
vortex states are unstable with respect to the Meissner state.

In conclusion, we have produced precise numerical solutions to the
GL equations consisting of infinite lattices of {}``doubly quantized''
vortices in bulk superconductors. The calculations can be carried
out efficiently for mean inductions down to 10\% of the upper critical
value. Although such solutions of the GL equations never globally
minimize the GL free energy for bulk superconductors, we expect they
will be useful as starting points for solving the GL equations in
film geometry.

\appendix

\section{\label{sec:AppendixA}Solving the Linearized GL equations in Terms
of the $a_{\mathbf{K}}$}

In his pioneering work on vortex lattices in superconductors, Abrikosov\cite{abrikosov_magnetic_1957}
showed that for an applied field just below $H_{c2}$, the first GL
equation (when expressed as an equation for the order parameter) has
the form of Schrödinger's equation for a charged particle confined
to a plane and subject to a magnetic field. With an assumed periodicity
of the vortex lines and one flux quantum per vortex, an analytic solution
$\psi_{A}$ exists and can be expressed\cite{abrikosov_magnetic_1957,kleiner_bulk_1964,brandt_die_1969}
in terms of a Jacobi theta function, 

\begin{equation}
\psi_{A}\left(x,y\right)=e^{-\pi y^{2}/x_{1}y_{2}}\vartheta_{1}\left(\frac{\pi}{x_{1}}\left(x+iy\right),\frac{x_{2}+iy_{2}}{x_{1}}\right)\label{eq:JacobiAnsatz}\end{equation}
where $\vartheta_{1}\left(z,\tau\right)\equiv2\sum_{n=0}^{\infty}\left(-\right)^{n}e^{i\pi\tau\left(n+\frac{1}{2}\right)^{2}}\sin(2n+1)z$
and the lattice parameters $x_{1}$, $x_{2}$ and $y_{2}$ were defined
just below Eq.~\eqref{eq:qansatz}. With this form for $\psi_{A}$
its modulus squared, $\omega_{A}\equiv\left|\psi_{A}\right|^{2}$,
is expressed as a double sum. This leads\cite{brandt_ginsburg-landau_1972}
to the Fourier like expansion of real terms, $\omega_{A}=\sum_{\mathbf{K}}a_{\mathbf{K}}^{A}\left[1-\cos\left(\mathbf{K}\cdot\mathbf{r}\right)\right]$,
with $a_{\mathbf{K}}^{A}=-\left(-\right)^{m+mn+n}e^{-K_{mn}^{2}S/8\pi}$.
Note that the sum over $\mathbf{K}$ is still a double sum over $m$
and $n$. 

Lasher\cite{lasher_series_1965} pointed out that for vortices of
multiplicity $p$, $\psi_{A}^{\left(p\right)}\left(r\right)=\left[\psi_{A}\left(r/\sqrt{p}\right)\right]^{p}$
is a corresponding solution of the linearized GL equations. In principle
one could use this form to determine $a_{\mathbf{K}}^{A}$ for doubles
in the expansion \eqref{eq:omegaansatz1}, starting from \eqref{eq:JacobiAnsatz},
but we did not attempt to carry that through.

We have taken an alternative approach based on numerical solution
of a linear system for the $a_{\mathbf{K}}^{A}$ derived from the
linearized GL equations. In the linear regime $b\left(\mathbf{r}\right)=B(0)-\bar{B}-\omega_{A}\left(\mathbf{r}\right)/2\kappa$
(see, for example, De~Gennes\cite{gennes_superconductivity_1966}
Sec.~6.7). Taking the curl and combining with the second GL equation
yields \begin{equation}
\frac{1}{2\kappa}\nabla\omega_{A}\times\hat{\mathbf{z}}=\omega_{A}\mathbf{Q}_{A}\label{eq:linearizedfirstGL}\end{equation}
Combining \eqref{eq:linearizedfirstGL}, \eqref{eq:omegaansatz2},
and the Fourier expansion for the supervelocity in the Abrikosov limit
(see Eq.~(24) in Ref.~\onlinecite{brandt_ginsburg-landau_1972}),
then applying uniqueness of Fourier series leads to the linear system

\begin{equation}
\sum_{\mathbf{K}_{i}}A_{ji}a_{\mathbf{K}_{i}}^{A}=-\langle\omega_{A}\rangle\label{eq:aKA_LinSol}\end{equation}
with

\begin{equation}
A_{ji}\equiv C_{ji}-\delta_{\mathbf{K}_{i},\mathbf{K}_{j}}\frac{1}{2\bar{B}\mathbf{\kappa}}\left|\mathbf{K}_{j}\right|^{2}+\delta_{\mathbf{K}_{i},\mathbf{K}_{j}/2}\frac{1}{8\bar{B}\mathbf{\kappa}}\left|\mathbf{K}_{j}\right|^{2}\label{eq:DesignMatrix1}\end{equation}
and

\begin{equation}
\begin{aligned}C_{ji}\equiv-\frac{\left|\mathbf{K}_{j}\right|^{2}-\mathbf{K}_{i}\cdot\mathbf{K}_{j}}{\left|\mathbf{K}_{j}-\mathbf{K}_{i}\right|^{2}}-\frac{\left|\mathbf{K}_{j}\right|^{2}+\mathbf{K}_{i}\cdot\mathbf{K}_{j}}{\left|\mathbf{K}_{j}+\mathbf{K}_{i}\right|^{2}}\\
+\hspace{3pt}\frac{\left|\mathbf{K}_{j}\right|^{2}-2\mathbf{K}_{i}\cdot\mathbf{K}_{j}}{4\left|\mathbf{K}_{j}-2\mathbf{K}_{i}\right|^{2}}+\frac{\left|\mathbf{K}_{j}\right|^{2}+2\mathbf{K}_{i}\cdot\mathbf{K}_{j}}{4\left|\mathbf{K}_{j}+2\mathbf{K}_{i}\right|^{2}}\end{aligned}
\label{eq:DesignMatrix2}\end{equation}
We follow Brandt's convention that $\langle\omega_{A}\rangle=1$,
so $(3/2)\sum_{\mathbf{K}}a_{\mathbf{K}}^{A}=1$. 

The infinite system of equations \eqref{eq:aKA_LinSol} could be rendered
finite by setting $a_{\mathbf{K}}^{A}=0$ for $|\mathbf{K}|>K_{\mathrm{max}}$;
however, this is not a good closure assumption because of slow convergence
with increasing $K_{\mathrm{max}}$. Eq.~\eqref{eq:DesignMatrix1}
shows the strong connection between $a_{\mathbf{K}}^{A}$ and $a_{\mathbf{K}/2}^{A}$
previously mentioned in Sec.~\ref{sub:Implementation-issues}. In
particular, for $K_{\mathrm{max}}/2<|\mathbf{K}|\leq K_{\mathrm{max}}$
the corresponding $a_{\mathbf{K}}^{A}$ are connected to coefficients
associated with vectors beyond the cutoff. In addition, as $|\mathbf{K}_{j}|\to\infty$,
$C_{ji}\to-3/2$, which leads to $a_{\mathbf{K}}^{A}\approx a_{\mathbf{K}/2}^{A}/4$
at large $|\mathbf{K}|$. (For large $|\mathbf{K}_{j}|$, $\sum_{\mathbf{K}_{i}}C_{ji}a_{\mathbf{K}_{i}}^{A}\approx-(3/2)\sum_{\mathbf{K}}a_{\mathbf{K}}^{A}=-\langle\omega_{A}\rangle$,
hence the latter two terms on the right side of Eq.~\ref{eq:DesignMatrix1}
must sum to zero.) We therefore set $a_{2\mathbf{K}}^{A}=a_{\mathbf{K}}^{A}/4$,
$a_{4\mathbf{K}}^{A}=a_{\mathbf{K}}^{A}/16$, and so on for $K_{\mathrm{max}}/2<|\mathbf{K}|\leq K_{\mathrm{max}}$
and this leads to a modified linear system with coefficients $A'_{ji}$.
For $i$ such that $K_{\mathrm{max}}/2<|\mathbf{K}_{i}|\leq K_{\mathrm{max}}$,
\begin{equation}
A'_{ji}=A_{ji}+\sum_{l=1}^{\infty}4^{-l}C_{j,2^{l}i}\label{eq:modifiedAmatrix}\end{equation}
 with $\mathbf{K}_{2^{l}i}\equiv2^{l}\mathbf{K}_{i}$. We truncate
the sum at $l=4$ after finding no change in the results for $a_{\mathbf{K}}^{A}$
when further terms are included. 

\begin{figure}
\includegraphics[width=1\columnwidth]{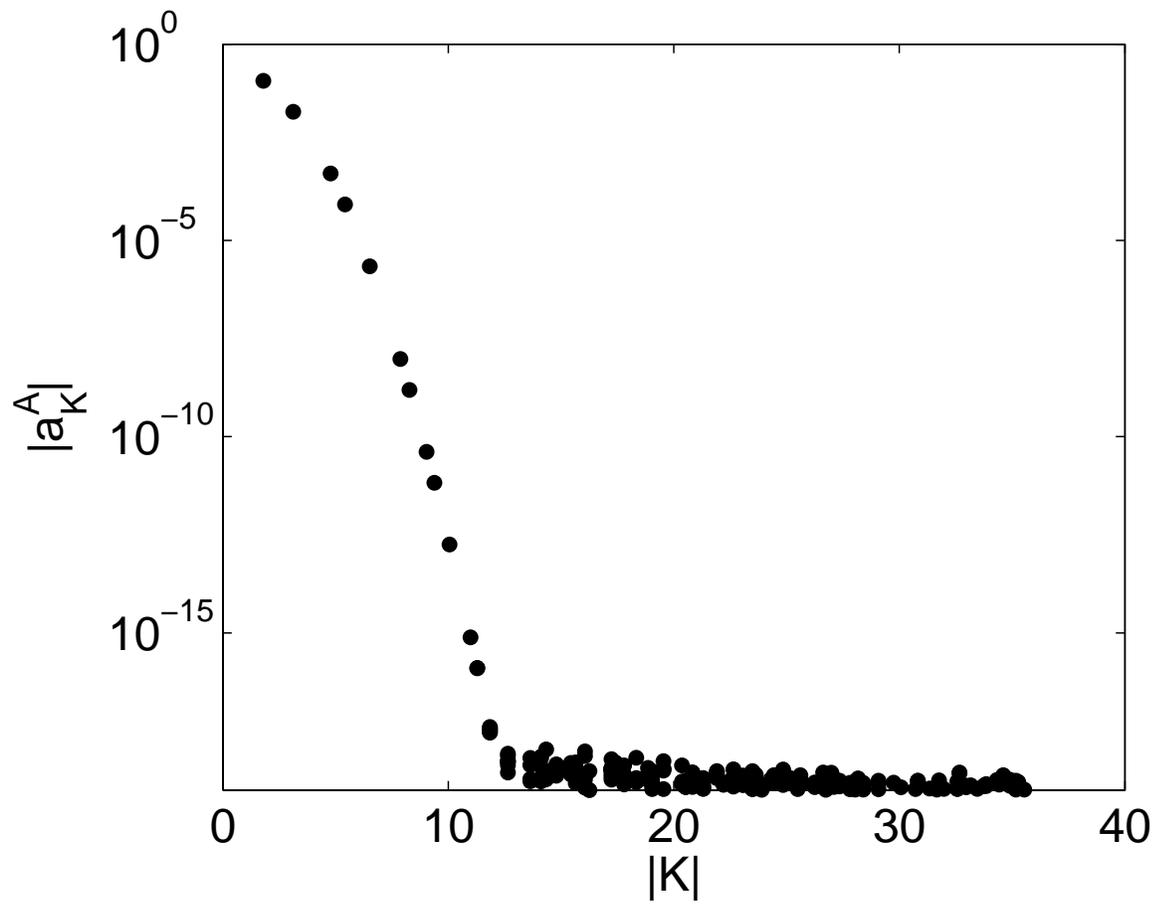}\caption{\label{fig:Fundamentals}Solutions to the linear system for the $a_{\mathbf{K}}^{A}$
associated with fundamental reciprocal lattice vectors.}

\end{figure}

\begin{figure}
\includegraphics[width=1\columnwidth]{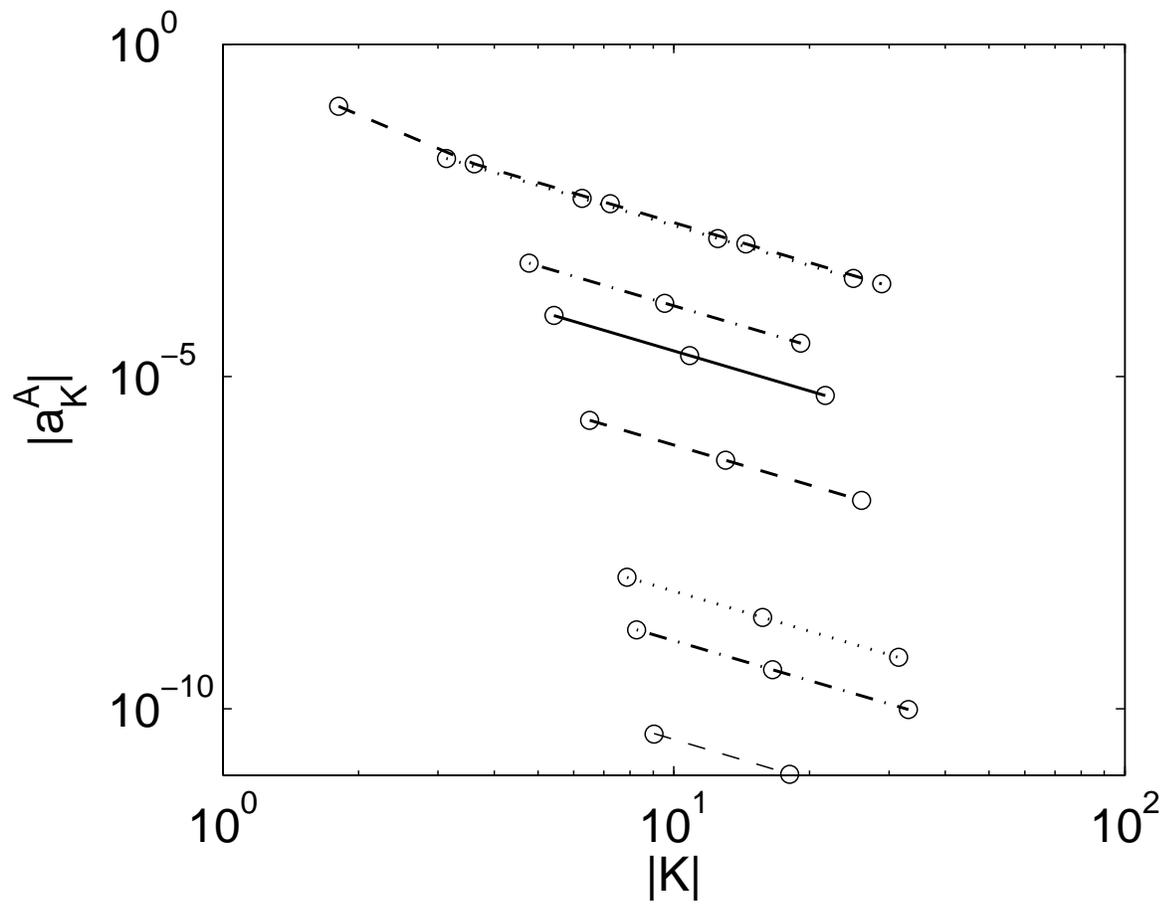}\caption{\label{fig:aKA_families}Solutions to the linear system for the $a_{\mathbf{K}}^{A}$
associated with eight fundamental reciprocal lattice vectors and those
vectors multiplied by powers of two. The lines are guides to the eye
connecting $a_{\mathbf{K}}^{A}$, $a_{2\mathbf{K}}^{A}$, $a_{4\mathbf{K}}^{A}$,\ldots{}}

\end{figure}

In Figs.~\ref{fig:Fundamentals} and \ref{fig:aKA_families} we show
the results of numerically solving the linear system for $\kappa=1$
and $K_{\mathrm{max}}=36$. In the former only $a_{\mathbf{K}}^{A}$
for fundamental reciprocal lattice vectors are shown; while in the
latter $a_{\mathbf{K}}^{A}$ for reciprocal lattice vectors that are
powers of two times several different fundamentals are displayed,
showing that $a_{\mathbf{K}}^{A}\approx a_{\mathbf{K}/2}^{A}/4$ holds
even for $|\mathbf{K}|$ not very large.

Finally, let us note that as a check on this approach to solving the
linearized GL equations we have carried an analogous analysis for
singles. The numerical results from solving the corresponding linear
equation for the $a_{\mathbf{K}}^{A}$ match the exact, analytic results. 

\bibliographystyle{apsrev}

\begin{thebibliography}{15}
\expandafter\ifx\csname natexlab\endcsname\relax\def\natexlab#1{#1}\fi
\expandafter\ifx\csname bibnamefont\endcsname\relax
  \def\bibnamefont#1{#1}\fi
\expandafter\ifx\csname bibfnamefont\endcsname\relax
  \def\bibfnamefont#1{#1}\fi
\expandafter\ifx\csname citenamefont\endcsname\relax
  \def\citenamefont#1{#1}\fi
\expandafter\ifx\csname url\endcsname\relax
  \def\url#1{\texttt{#1}}\fi
\expandafter\ifx\csname urlprefix\endcsname\relax\def\urlprefix{URL }\fi
\providecommand{\bibinfo}[2]{#2}
\providecommand{\eprint}[2][]{\url{#2}}

\bibitem[{\citenamefont{Abrikosov}(1957)}]{abrikosov_magnetic_1957}
\bibinfo{author}{\bibfnamefont{A.~A.} \bibnamefont{Abrikosov}},
  \bibinfo{journal}{Zh. Eksp. Teor. Fiz.} \textbf{\bibinfo{volume}{32}},
  \bibinfo{pages}{1442} (\bibinfo{year}{1957}).

\bibitem[{\citenamefont{Matricon}(1966)}]{matricon_thesis}
\bibinfo{author}{\bibfnamefont{J.}~\bibnamefont{Matricon}}, Ph.D. thesis,
  \bibinfo{school}{Universit\'e de Paris} (\bibinfo{year}{1966}).

\bibitem[{\citenamefont{Lasher}(1967)}]{lasher_mixed_1967}
\bibinfo{author}{\bibfnamefont{G.}~\bibnamefont{Lasher}},
  \bibinfo{journal}{Phys. Rev.} \textbf{\bibinfo{volume}{154}},
  \bibinfo{pages}{345} (\bibinfo{year}{1967}).

\bibitem[{\citenamefont{Callaway}(1992)}]{callaway_magnetic_1992}
\bibinfo{author}{\bibfnamefont{D.~J.~E.} \bibnamefont{Callaway}},
  \bibinfo{journal}{Ann. Phys. (NY)} \textbf{\bibinfo{volume}{213}},
  \bibinfo{pages}{166} (\bibinfo{year}{1992}).

\bibitem[{\citenamefont{Hasegawa et~al.}(1991)\citenamefont{Hasegawa, Matsuda,
  Endo, Osakabe, Igarashi, Kobayashi, Naito, Tonomura, and
  Aoki}}]{hasegawa_magnetic-flux_1991}
\bibinfo{author}{\bibfnamefont{S.}~\bibnamefont{Hasegawa}},
  \bibinfo{author}{\bibfnamefont{T.}~\bibnamefont{Matsuda}},
  \bibinfo{author}{\bibfnamefont{J.}~\bibnamefont{Endo}},
  \bibinfo{author}{\bibfnamefont{N.}~\bibnamefont{Osakabe}},
  \bibinfo{author}{\bibfnamefont{M.}~\bibnamefont{Igarashi}},
  \bibinfo{author}{\bibfnamefont{T.}~\bibnamefont{Kobayashi}},
  \bibinfo{author}{\bibfnamefont{M.}~\bibnamefont{Naito}},
  \bibinfo{author}{\bibfnamefont{A.}~\bibnamefont{Tonomura}}, \bibnamefont{and}
  \bibinfo{author}{\bibfnamefont{R.}~\bibnamefont{Aoki}},
  \bibinfo{journal}{Phys. Rev. B} \textbf{\bibinfo{volume}{43}},
  \bibinfo{pages}{7631} (\bibinfo{year}{1991}).

\bibitem[{\citenamefont{Sweeney and Gelfand}(2010)}]{sweeney_simple_2010}
\bibinfo{author}{\bibfnamefont{M.~C.} \bibnamefont{Sweeney}} \bibnamefont{and}
  \bibinfo{author}{\bibfnamefont{M.~P.} \bibnamefont{Gelfand}},
  \bibinfo{journal}{arXiv:1003.0648v1}  (\bibinfo{year}{2010}).

\bibitem[{\citenamefont{Brandt}(1972)}]{brandt_ginsburg-landau_1972}
\bibinfo{author}{\bibfnamefont{E.~H.} \bibnamefont{Brandt}},
  \bibinfo{journal}{Phys. Status Solidi B} \textbf{\bibinfo{volume}{51}},
  \bibinfo{pages}{345} (\bibinfo{year}{1972}).

\bibitem[{\citenamefont{Brandt}(1997)}]{brandt_precision_1997}
\bibinfo{author}{\bibfnamefont{E.~H.} \bibnamefont{Brandt}},
  \bibinfo{journal}{Phys. Rev. Lett.} \textbf{\bibinfo{volume}{78}},
  \bibinfo{pages}{2208} (\bibinfo{year}{1997}).

\bibitem[{\citenamefont{Tinkham}(1996)}]{tinkham_introduction_1996}
\bibinfo{author}{\bibfnamefont{M.}~\bibnamefont{Tinkham}},
  \emph{\bibinfo{title}{Introduction to superconductivity}}
  (\bibinfo{publisher}{{McGraw} Hill}, \bibinfo{address}{New York},
  \bibinfo{year}{1996}), \bibinfo{edition}{2nd} ed.

\bibitem[{\citenamefont{Klein and P\"ottinger}(1991)}]{klein_virial_1991}
\bibinfo{author}{\bibfnamefont{U.}~\bibnamefont{Klein}} \bibnamefont{and}
  \bibinfo{author}{\bibfnamefont{B.}~\bibnamefont{P\"ottinger}},
  \bibinfo{journal}{Phys. Rev. B} \textbf{\bibinfo{volume}{44}},
  \bibinfo{pages}{7704} (\bibinfo{year}{1991}).

\bibitem[{\citenamefont{Doria et~al.}(1989)\citenamefont{Doria, Gubernatis, and
  Rainer}}]{doria_virial_1989}
\bibinfo{author}{\bibfnamefont{M.~M.} \bibnamefont{Doria}},
  \bibinfo{author}{\bibfnamefont{J.~E.} \bibnamefont{Gubernatis}},
  \bibnamefont{and} \bibinfo{author}{\bibfnamefont{D.}~\bibnamefont{Rainer}},
  \bibinfo{journal}{Phys. Rev. B} \textbf{\bibinfo{volume}{39}},
  \bibinfo{pages}{9573} (\bibinfo{year}{1989}).

\bibitem[{\citenamefont{Kleiner et~al.}(1964)\citenamefont{Kleiner, Roth, and
  Autler}}]{kleiner_bulk_1964}
\bibinfo{author}{\bibfnamefont{W.~H.} \bibnamefont{Kleiner}},
  \bibinfo{author}{\bibfnamefont{L.~M.} \bibnamefont{Roth}}, \bibnamefont{and}
  \bibinfo{author}{\bibfnamefont{S.~H.} \bibnamefont{Autler}},
  \bibinfo{journal}{Phys. Rev.} \textbf{\bibinfo{volume}{133}},
  \bibinfo{pages}{A1226} (\bibinfo{year}{1964}).

\bibitem[{\citenamefont{Brandt}(1969)}]{brandt_die_1969}
\bibinfo{author}{\bibfnamefont{E.~H.} \bibnamefont{Brandt}},
  \bibinfo{journal}{Phys. Status Solidi B} \textbf{\bibinfo{volume}{36}},
  \bibinfo{pages}{393} (\bibinfo{year}{1969}).

\bibitem[{\citenamefont{Lasher}(1965)}]{lasher_series_1965}
\bibinfo{author}{\bibfnamefont{G.}~\bibnamefont{Lasher}},
  \bibinfo{journal}{Phys. Rev.} \textbf{\bibinfo{volume}{140}},
  \bibinfo{pages}{A523} (\bibinfo{year}{1965}).

\bibitem[{\citenamefont{de~Gennes}(1966)}]{gennes_superconductivity_1966}
\bibinfo{author}{\bibfnamefont{P.}~\bibnamefont{de~Gennes}},
  \emph{\bibinfo{title}{Superconductivity of metals and alloys}}
  (\bibinfo{publisher}{{W.A.} Benjamin}, \bibinfo{address}{New York},
  \bibinfo{year}{1966}).

\end{thebibliography}


\end{document}